\documentclass{iaus}
\usepackage{graphicx}

\title[Review about populations of Be stars: stellar evolution of extreme stars] %% give here short title %%
{Review about populations of Be stars: stellar evolution of extreme stars}

\author[C. Martayan, T. Rivinius, D. Baade, A.-M. Hubert, \& J. Zorec]   %% give here short author list %%
{C. Martayan$^{1,2}$, T. Rivinius$^1$, D. Baade$^3$, A.-M. Hubert$^2$,\\ 
\and J. Zorec$^4$}

\affiliation{$^1$ESO Chile   \\[\affilskip]
$^2$GEPI-Observatoire de Meudon, France    \\[\affilskip]
$^3$ESO Germany   \\[\affilskip]
$^4$Institut d'Astrophysique de Paris, France
\\email: {\tt cmartaya@eso.org}
}

\pubyear{2011}
\volume{272}  %% insert here IAU Symposium No.
\pagerange{1--12}
\jname{Active OB stars: structure, evolution, mass loss and critical limits}
\editors{C.\ Neiner, G.\ Wade, G.\ Meynet \& G.\ Peters, eds.}
\begin{document}

\maketitle

\begin{abstract}
Among the emission-line stars, the classical Be stars known for their extreme properties are remarkable. 
The Be stars are B-type main sequence stars that have displayed at least once in their life emission lines in their spectrum. 
Beyond this phenomenological approach some progresses were made on the understanding of this class of stars. 
With high-technology techniques (interferometry, adaptive optics, multi-objects spectroscopy, spectropolarimetry, 
high-resolution photometry, etc) from different instruments and space mission such as the VLTI, CHARA, FLAMES, ESPADONS-NARVAL, 
COROT, MOST, SPITZER, etc, some discoveries were performed allowing to constrain the modeling of the Be stars stellar evolution 
but also their circumstellar decretion disks. In particular, the confrontation between theory and observations about the effects 
of the stellar formation and evolution on the main sequence, the metallicity, the magnetic fields, the stellar pulsations, 
the rotational velocity, and the binarity (including the X-rays binaries) on the Be phenomenon appearance is
discussed. The disks observations and the efforts made on their modeling is mentioned. 
As the life of a star does not finish at the end of the main sequence, we also mention their stellar evolution post main sequence 
including the gamma-ray bursts.
Finally, the different new results and remaining questions about the main physical properties of the Be stars are summarized
and possible ways of investigations proposed. The recent and future facilities (XSHOOTER, ALMA, E-ELT, TMT, GMT, JWST, GAIA, etc) and
their instruments that may help to improve the knowledge of Be stars are also briefly introduced.

\end{abstract}

\firstsection % if your document starts with a section,
\section{Introduction}

The emission-line stars are spread over the entire HR digram. They concern young, not evolved stars or at the opposite evolved
stars and of course Main Sequence stars. They are or not massive stars.
For instance, among others, one can found  TTauri stars, UV Ceti stars, flare stars, Mira stars, HBe/Ae stars, 
B[e] stars, Be/Oe stars, Of stars, Supergiant stars, LBV, WR stars.
This document concerns mainly OeBeAe stars. 
The OeBeAe stars have properties similar to OBA type star (see Evans contribution, this volume) and specific properties
due to their extreme nature.
The first Be star, $\gamma$ Cas was discovered by the father Secchi in 1867 (\cite[Secchi 1967]{secchi1867}).

\begin{figure}[h]
% \vspace*{-2.0 cm}
\begin{center}
 \includegraphics[width=3.4in]{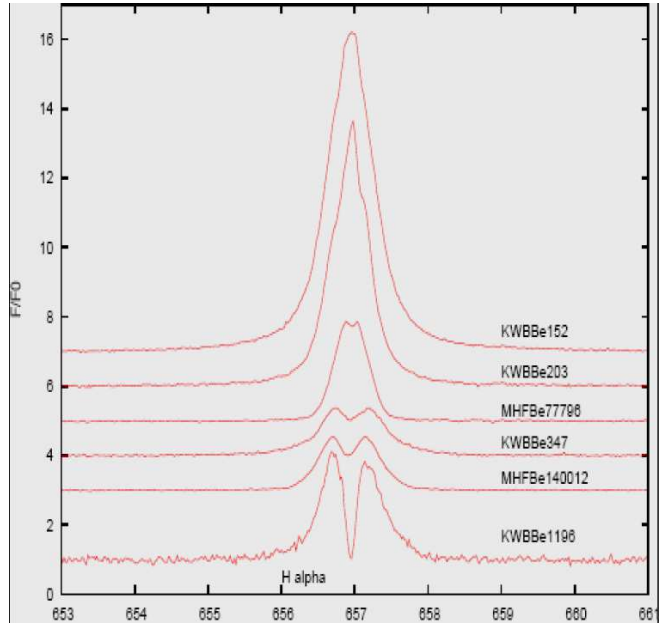} 
% \vspace*{-1.0 cm}
 \caption{Examples of H$\alpha$ emission-line in different LMC Be stars. 
 The shape depends among other parameter of the inclination angle.}
   \label{fig1}
\end{center}
\end{figure}

\cite[Collins (1987)]{collins1987} wrote the first definition of Be stars:
``they are non-supergiant OBA-type stars that have displayed at least once in their spectrum emission lines (H$\alpha$).''
As example, see Fig.~\ref{fig1}.
 Actually the emission-lines come from the circumstellar decretion disk (\cite[Struve 1931]{struve1931}) formed by episodic matter
 ejections from the central star.
 Are the matter ejections related to the rotation?
 Be stars are very fast rotators, rotating close to the breakup velocity but often an additional mechanism is needed
 for the star being able to eject the matter.
 In the next sections, the rotation and other mechanisms (binarity, stellar evolution, magnetic fields, pulsations) are examined as well as the effects of the metallicity, the stellar
 evolution, etc.

\cite[Porter \& Rivinius (2003)]{Porter2003} presented different kind of emission-line stars with their known properties
and how they differ to the ``classical Be stars''.
Since 2003, with the improvements of the technology (interferometry, MOS, etc) and the efforts made on the modeling and theory,
the knowledge of Be stars was improved. The next sections will provide some very summarized informations (since 2004, more than 240
refeered articles were published dealing with Be stars) about these recent developments (see also contribution by Baade, this volume).

{\underline{\it -Techniques for detecting a Be star}}

Before studying Be stars, it is necessary to find/recognize them.
There are several ways to detect them.
The first possibility is to use the photometric techniques.
Combining different colour/colour or colour/magnitude diagrams would help to pre-classify the stars and
detect potential Be star candidates.
\cite[Keller et al. (1999)]{keller99} provide examples of CMD with given thresholds above which the Be stars could fall.
They also show that Be stars tend to form a redder sequence than normal B stars.%, see Fig.~\ref{fig2}.

\cite[Dachs et al. (1988)]{dachs1988} have shown that the infrared excess is related to the circumstellar disk of Be stars.
It is also linked to the H$\alpha$ equivalent width.
More recent studies found similar results in other wavelength domains such as in the infrared with the AKARI survey
(\cite[Ita et al. 2010]{ita2010}) or with SPITZER (\cite[Bonanos et al. 2010]{bonanos2010}).
However, this kind of study has some limits due to the intrinsic properties of Be stars (variability, change of phases Be-B).
\cite[McSwain \& Gies (2005)]{McSwaingies2005} in their photometric survey of Galactic open clusters found 63 \% 
of the known Be stars due to the transience of the Be phenomenon.
Thus with the photometry the coupling of indices is necessary.
It is also necessary to disentangle the local reddening and the infrared excess 
due to the circumstellar disk of Be stars.

Another way to find the emission-line stars, and the Be stars, is to do slitless spectroscopy and combine the spectroscopic diagnostic
with the photometry as presented in \cite[Martayan et al. (2010a)]{marta2010a}.
The slitless spectroscopy gives the advantage to do not be sensitive to the diffuse nebulosity and the spectra are not contaminated
by other nebular lines.

Another possibility for finding Be stars is to study the behaviour and the variation of lightcurves.
The Be phenomenon is transient, as a consequence the star could appear as a Be star or as a B star when the disk was blown up.
Moreover, Be stars show often outbursts.
These characteristics help to find this kind of stars using the lightcurves variations as published by
\cite[Mennickent et al. (2002)]{mennickent2002} and \cite[Sabogal et al. (2005)]{sabogal2005}.

{\underline{\it -The Surveys}}

Be stars with the techniques mentionned above can be found mainly in open clusters and in fields.
In the Galaxy, there are different surveys, let us mention, the photometric survey by 
\cite[McSwain \& Gies (2005)]{McSwaingies2005} and the slitless spectroscopic survey by 
\cite[Mathew et al. (2008)]{Mathew2008}.

\begin{figure}[h]
% \vspace*{-2.0 cm}
\begin{center}
 \includegraphics[width=3.4in]{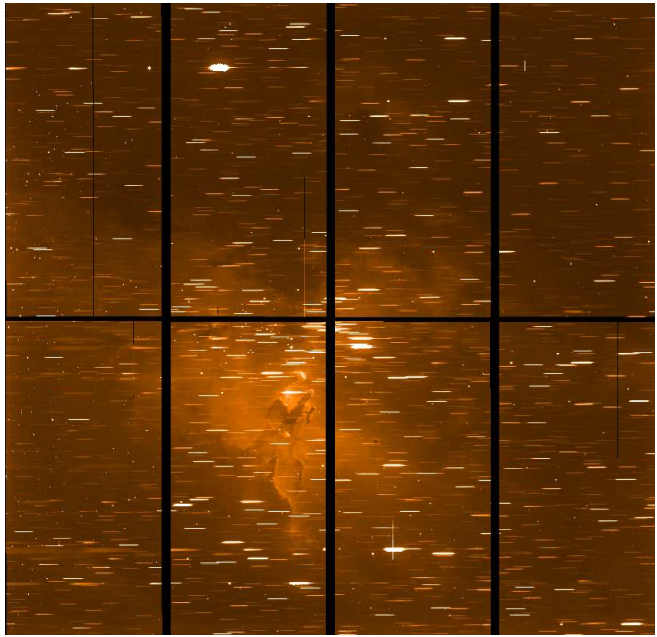} 
% \vspace*{-1.0 cm}
 \caption{Example of slitless spectroscopy in the NGC6611 open cluster by \cite[Martayan et al. (2008)]{Marta2008}.
 The stars appear as spectra.}
   \label{fig4}
\end{center}
\end{figure}

In other galaxies such as the Magellanic Clouds but also in the Milky Way, there are the following surveys
among others:
a slitless spectroscopic survey of the SMC by \cite[Meyssonnier \& Azzopardi (1993)]{MA1993},
another in the SMC/LMC/MW by \cite[Martayan et al. (2010a)]{Marta2010a}, \cite[Martayan et al. (2008)]{Marta2008} 
(see Fig.~\ref{fig4}).

And recently, with the VLT-FLAMES, some spectroscopic surveys were and are performed in different galaxies 
(\cite[Evans et al. 2005]{evans2005}, \cite[Martayan et al. 2006a]{Marta2006a}, 
\cite[Martayan et al. 2007a]{Marta2007a}).
With the polarimetry, 
some Be stars were found in the SMC/LMC (\cite[Wisniewski et al. 2007]{wis2007}, 
\cite[Wisniewski \& Bjorkman 2006]{wis2006}),   
and with FORS2 in IC1613 (\cite[Bresolin et al. 2007]{bresolin2007}), which correspond to the farthest Be stars detected.

\section{Mid/long term monitoring of Be stars, V/R variations}

{\underline{\it -Monitoring of Be stars}}

Why long-term monitorings are needed?
Due to the transience of the Be phenomenon for having a chance to detect a Be star, a mid or long term
monitoring of B/Be stars is needed. Indeed, for the follow-up of the variability of the star 
related to outbursts and/or pulsations, or binarity, or modifications of the CS disk, etc, a monitoring is needed.

Surveys such as the photometric ones of MACHO or OGLE are useful in that sense 
but also the spectroscopic surveys.
Here the amateur astronomers have a role to play. With the new efficient, good quality, low cost
spectrograph that can be mounted in amateur telescopes, see contribution by Blanchard et al. (this volume), some useful good quality
observations can be performed.

Moreover, without long-term monitoring (more than 10 years of spectroscopy) 
the results concerning the Be star $\mu$ Cen and the predictions
of the outbursts dates could not have been obtained.
For all details about this pioneer study, see \cite[Rivinius et al. (1998a)]{rivi1998a}, 
\cite[Rivinius et al. (1998b)]{rivi1998b}.

This kind of study could thanks to the databases be performed more easily now, let us cite the 
Be Star Spectra database http://basebe.obspm.fr/basebe/
(\cite[Neiner et al. 2007]{neiner2007}, contribution by De Batz et al., this volume). 

{\underline{\it -The V/R variation of the emission lines}}

\begin{figure}[h]
% \vspace*{-2.0 cm}
\begin{center}
 \includegraphics[width=5in]{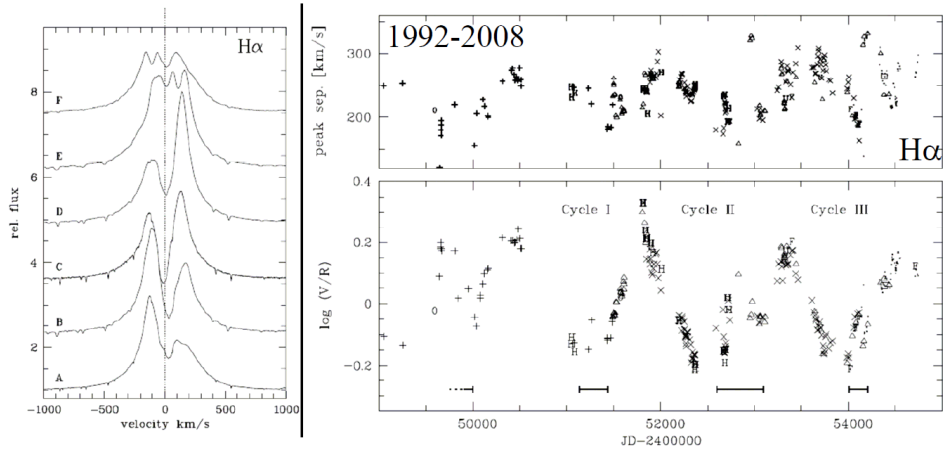} 
% \vspace*{-1.0 cm}
 \caption{Example of V/R variations in the Be star $\zeta$ Tau. Figure adapted from \cite[Stefl et al. (2009)]{stefl2009}.}
   \label{fig3b}
\end{center}
\end{figure}

A long-term spectroscopic monitoring of Be stars can show in certain cases variations 
of the V and R peaks of the emission-lines. This variation is now understood and is related to global 
one-armed oscillations in the circumstellar disk.
The case of the Be star $\zeta$ Tau (see \cite[Stefl et al. 2009]{stefl2009}) is representative
of that phenomenon with a V/R cycle of  about 1400 days without correlation with the period of 133 days
of the companion. Fig.~\ref{fig3b} from \cite[Stefl et al. (2009)]{stefl2009} illustrates this phenomenon.
The (polarimetric, spectroscopic, interferometric) observations are properly reproduced by a modeling 
of a one-armed spiral viscous disk as described in 
\cite[Carciofi et al. (2009)]{carciofi2009}, see also contributions by Carciofi and Stee, this volume.

\section{Stellar winds}

The diagnostic lines are mostly present in the UV and optical but also several of them are
in the infrared. Let us mention the lines of HeI, HeII, Si IV, C IV, N V, Br$\gamma$, etc.
For more details, see \cite[Martins et al. (2010)]{martins2010}, 
\cite[Mokiem et al. (2006)]{mokiem2006}, and Henrichs et al., this volume. 

The structure of the wind differs among B stars.
For the normal B stars, the wind is spherical, while for Be stars, we can expect a radiative bi-polar wind
mainly due to the Von Zeipel effect. The poles in case of fast rotation becomes brighter than 
the equator, it explains this asymetry.
In case of Be stars, one can also expect an additional mechanical wind
at the equator that creates the CS disk. In such case the needed value of the $\Omega/\Omega$$_{c}$
ratio (angular velocity to critical angular velocity ratio) for ejecting matter has to be determined (see Sect.\ \ref{rotvel}).

Fortunately, recent interferometric facilities (such as the VLTI-AMBER), allowed to measure the shape
of the CS environment of Be stars and both winds were found.
It is for example the case of the Be stars $\alpha$ Arae, Achernar 
(\cite[Meilland et al. 2007]{meilland2007}, \cite[Kervella et al. 2006]{kervella2006}).

\section{Rotational velocities}
\label{rotvel}

\begin{figure}[h!]
% \vspace*{-2.0 cm}
\begin{center}
 \includegraphics[width=3.4in]{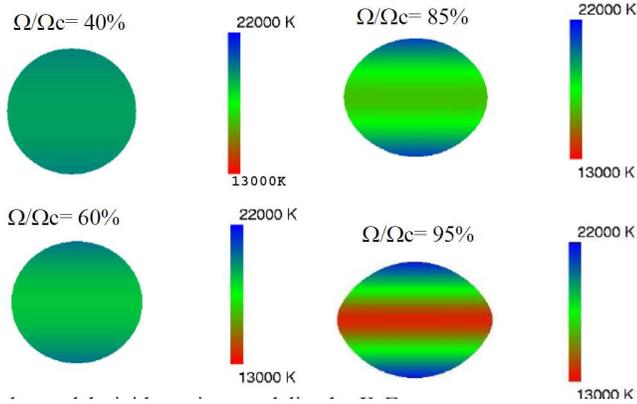} 
% \vspace*{-1.0 cm}
 \caption{Example of star shape depending on the $\Omega/\Omega$$_{c}$ ratio in the Roche model and rigid rotation
 approach. Due to the Von Zeipel effect there is a temperature gradient from the poles to the equator. 
 Courtesy by Y. Fr\'emat, see also \cite[Fr\'emat et al. (2005)]{fremat2005}.}
   \label{figdeforme}
\end{center}
\end{figure}

Be stars are known to be very fast rotator but how fast are they rotating?
Fig.~\ref{figdeforme} shows effect of the rotation on the shape of the stars
for different $\Omega/\Omega$$_{c}$ in the Roche model, rigid rotation approach.
Next sections provide some clues about this point.

{\underline{\it -From the interferometry}}

The interferometric observations show that all observed Be stars are found flattened 
revealing high $\Omega/\Omega$$_{c}$ratios.
The flattening of Achernar was found equal to 1.56 $\pm$0.05 by \cite[Domiciano de Souza et al. (2003)]{domi2003}.
However, this value could also be due to remaining small disk indicating that Achernar is not rotating
at the breakup velocity but close to it.

{\underline{\it -Line saturation effect at high $\Omega/\Omega$$_{c}$}}

\begin{figure}[h!]
% \vspace*{-2.0 cm}
\begin{center}
 \includegraphics[width=3.4in]{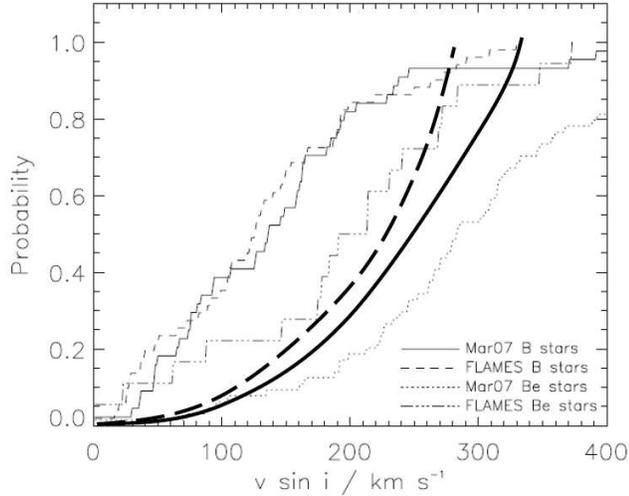} 
% \vspace*{-1.0 cm}
 \caption{Vsini measuremnts of B and Be stars by \cite[Hunter et al. (2008)]{hunter2008} compared to those 
 of \cite[Martayan et al. 2007a]{Marta2007a}. The dotted curve corresponds to $\Omega/\Omega$$_{c}$=70\% and the solid curve
 to $\Omega/\Omega$$_{c}$=80\% (line saturation effect with 1 He line measurement). 
 The curves come from \cite[McSwain et al. (2008)]{mcswain2008}.}
   \label{fig6}
\end{center}
\end{figure}

\cite[Townsend et al. (2004)]{townsend2004} showed that the HeI 4471 \AA \ line has a saturation effect at high
 $\Omega/\Omega$$_{c}$, typically above 80\%. This implies that the Vsini measurements based on the use of 1 He line 
 are limited in case of fast rotating stars, typically Be stars and possibly Bn stars.
 Using several He lines as well as metallic lines, the saturation effect occurs again but at higher  $\Omega/\Omega$$_{c}$
 ($\sim$90-95\%), see \cite[Fr\'emat et al. (2005)]{fremat2005}.
This explains the discrepancy between the measurements of Be stars Vsini between  
\cite[Hunter et al. (2008)]{hunter2008} and \cite[Martayan et al. (2007a)]{Marta2007a}.
Fig.~\ref{fig6} shows the measurements differences from \cite[Hunter et al. (2008)]{hunter2008} and the saturation curves at $\Omega/\Omega$$_{c}$=70 and 80 \% from
\cite[McSwain et al. (2008)]{mcswain2008}. It clearly shows that the Be star Vsini measurements done by 
 \cite[Hunter et al. (2008)]{hunter2008} are saturated and the measurements by \cite[Martayan et al. (2007a)]{Marta2007a}
 saturate at $\sim$90\%.
 
{\underline{\it -Fast rotation effects correction}}

Accordingly to the previous section, in case of fast rotators, one has to take into account the fast rotation effects 
(flattening, gravitational darkening). These effects introduce underestimates of the Vsini, and it implies too that
the spectral type/luminosity classes are modified from the late to earlier spectral type and from evolved to less
evolved luminosity classes. The codes BRUCE by \cite[Townsend et al. (2004)]{townsend2004} and
FASTROT by \cite[Fr\'emat et al. (2005)]{fremat2005} are able to properly correct the fundamental parameters of the fast
rotating stars.

{\underline{\it -The veiling effect}}

Another important effect in case of Be stars is the veiling effect due to the circumstellar disk 
that affects the continuum level and thus the lines depth. Not taking this effect into account will imply
bad measurements of the fundamental parameters but also of the chemical abundances.
For taking this effect into account, \cite[Ballereau et al. (1995)]{ballereau1995} proposed a method based on the He4471 emission-line EW
measurements.
Other possibility is to prefer the bluest lines when spectrum fitting is performed but also to determine the disk influence
by fitting the Spectral Energy Distribution. The new spectrocopic facility, the VLT-XSHOOTER that covers the near UV to the K-band
simultaneously should help in that approach.
Alternatively, there is the method consisting to fit the Balmer discontinuity as described by \cite[Zorec et al. (2009)]{zorec2009}

{\underline{\it -Metallicity effects}}

\cite[Keller (2004)]{keller2004}, \cite[Martayan et al. (2006a)]{marta2006a}, \cite[Martayan et al. (2007a)]{marta2007a}, 
and \cite[Hunter et al. (2008)]{hunter2008} found that SMC OB stars rotate faster than LMC OB stars, which rotate faster than
their Galactic counterparts.
In the SMC, the mass-loss of OB stars by radiatively driven winds was found lower than in the Galaxy by
\cite[Bouret et al. (2003)]{bouret2003} and \cite[Vink (2007)]{vink2007}.
Consequently, the stars should loose less angular momentum and could rotate faster (\cite[Maeder \& Meynet 2001]{MM2001}).
The comparisons between SMC Be stars Vsini fairly agree with the theoretical models by \cite[Ekstr\"om et al. (2008)]{ekstrom2008}.
The SMC Be stars are found to be rotating very close to the critical velocity.

{\underline{\it -ZAMS rotational velocities}}

The ZAMS rotational velocities of SMC, LMC, and Milky Way (MW) intermediate-mass Be stars were determined 
by \cite[Martayan et al. (2007a)]{marta2007a}.
They found a metallicity effect on the ZAMS rotational velocities. At lower metallicity Be stars rotate faster
since their birth. This could be related to an opacity effect, for an identical $\Omega/\Omega$$_{c}$,
at lower metallicitty, the radii are smaller thus the stars can rotate faster.
The comparison of the SMC ZAMS rotational velocities of intermediate-mass and massive Be stars with the theoretical tracks by 
\cite[Ekstr\"om et al. (2008)]{ekstrom2008} are also in a fair agreement.

% \begin{figure}[b]
%% \vspace*{-2.0 cm}
%\begin{center}
% \includegraphics[width=3.4in]{figBeZAMS.ps} 
%% \vspace*{-1.0 cm}
% \caption{ZAMS rotational velocities of Be stars samples by mass-categories in the SMC, LMC, 
% and MW from \cite[Martayan et al. (2007a)]{marta2007a}.}
%   \label{fig7}
%\end{center}
%\end{figure}

%\subsection{The differential rotation}

{\underline{\it -Number of Be stars vs. the metallicity and redshift}}

\begin{figure}[h!]
%% \vspace*{-2.0 cm}
\begin{center}
 \includegraphics[width=2.5in]{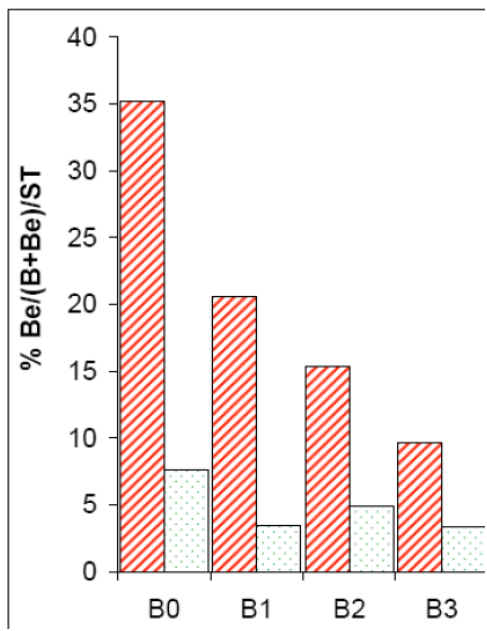} 
%% \vspace*{-1.0 cm}
 \caption{Ratio of Be to B stars by spectral categories in the SMC and MW. Figure adapted 
 from \cite[Martayan et al. (2010a)]{marta2010a}.Red dashed bars are for the SMC, white bars for the MW.}
   \label{fig8}
\end{center}
\end{figure}

As a consequence of faster rotational velocities at lower metallicity, one can expect to find more Be-type stars in low-metallicity
environments.
Using the ESO-WFI in its slitless mode (\cite[Baade et al. 1999]{baade1999}), \cite[Martayan et al. (2010a)]{marta2010a} found 3 to
5 times more Be stars in SMC open clusters than in the Galactic ones (\cite[McSwain \& Gies 2005]{McSwaingies2005}), 
see Fig.~\ref{fig8}.
This quantified the results by \cite[Maeder et al. (1999)]{maeder1999} and \cite[Wisniewski et al. (2008)]{wis2008}.

Another consequence of this increase of the Be stars number with decreasing metallicity is when a survey of OBA-type
populations is performed, one should find an increased number/ratio of OeBeAe stars in the sample.
This was exactly the case of the survey by \cite[Bresolin et al. (2007)]{bresolin2007} who found 6 Be stars in the 6 MS B stars
observed in the low metallicity galaxy IC1613.
Actually at lower metallicity than the SMC, and probably at higher redshift, the stars could rotate again faster, 
then one can expect to have more Be stars but also
that the Be phenomenon is extended to other categories of stars.

{\underline{\it -Chemical abundances}}

The measurement of chemical abundances is very important for testing the stellar evolution models with fast rotation, including the
rotational mixing due to the rotation. Moreover, it is also important for distinguishing the stars following a chemical homogeneous evolution than the usual
evolution (\cite[Maeder et al. 1987]{maeder1987}, \cite[Yoon et al. 2006]{yoon2006}).
It is also of interest to know them for the pulsating stars of different metallicity environments. 

\cite[Lennon et al. (2005)]{lennon2005} found 2 Be stars without N enhancement, while the theoretical models using the
rotational mixing expect a N enhancement and a C depletion.
Dunstall et al. (this volume) and Peters (this volume) also found no N enrichment in Be stars.
Does it mean that the rotational mixing theory is not able to explain the measurements?
Not necessarily, \cite[Porter (1999)]{porter1999} found that due to the temperature gradient in Be stars, the ions are displaced
from the poles to the equator. This could occur to the N, implying that the corresponding lines become weaker and the
measurements biased.

\section{Stellar pulsations}

Be stars lie in $\beta$ Cep and SPB regions, thus p and g modes are expected to be found. The $\kappa$ mechanism is at the origin of
the observed pulsations due to the iron bump.
A beating of non radial pulsations combined to the fast rotation of Be stars could be at the origin of the matter ejection.
According to  \cite[Townsend et al. (2004)]{townsend2004} an $\Omega/\Omega$$_{c}$=95\% is needed for launching the matter in orbit, while
\cite[Cranmer (2009)]{cranmer2009} shows that depending on the $\Omega/\Omega$$_{c}$ value the ejected matter will be part of the wind
or will form a disk.
\cite[Rivinius et al. (1998a)]{rivi1998a} have shown that the combination
fast rotation + beating of non radial pulsations is able to reproduce and predict the outbursts in the Be star $\mu$ Cen.
More recently, with the new photometric space missions MOST, COROT, KEPLER, a large number of pulsational frequencies 
(g and p modes) were found in several Be stars.
\cite[Huat et al. (2009)]{huat2009}, in the Be star HD49330, have found that the pulsational frequencies 
but also their amplitude change between the quiescence phase, the pre burst phase, and during the burst phase.
Is it related to the subsurface convection zones in OB stars? For more details, see Cantiello contribution (this volume).

 \begin{figure}[h!]
% \vspace*{-2.0 cm}
\begin{center}
 \includegraphics[width=4.5in]{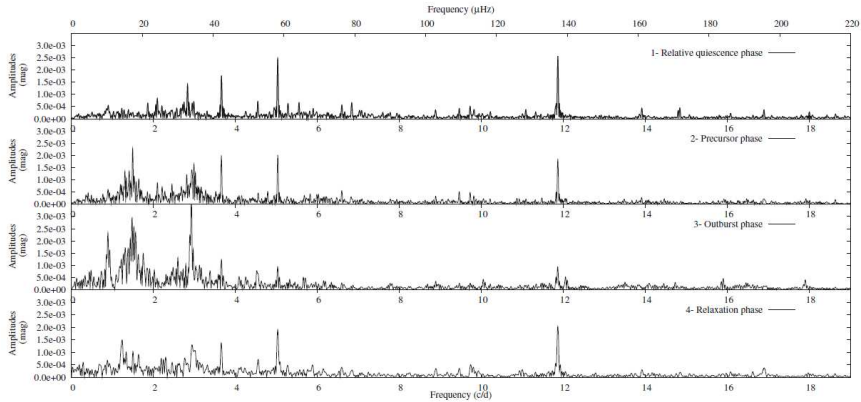} 
% \vspace*{-1.0 cm}
 \caption{Evolution of pulsational frequencies in the Be star HD49330 from COROT data during the different pre, during, and
 after an outburst phases. The figure comes from \cite[Huat et al. (2009)]{huat2009}.}
   \label{fig9}
\end{center}
\end{figure}

\section{Spectropolarimetry: magnetic fields and disks}
The spectropolarimetry is a useful tool for finding the Be stars due to their disks but also for detecting the magnetic fields.
Recent spectropolarimetric facilities such as ESPADONS, NARVAL, allow the search of magnetic fields in Be stars but up to now,
there is only 1 Be star in which a magnetic field was found: $\omega$ Ori (\cite[Neiner et al. 2003]{neiner2003}).
The presence of magnetic field ($\Omega$ transport from the core to the surface or magnetic reconnection) 
combined to the fast rotation could also be an additional mechanism for ejecting matter and create the Be star CS disk. 

\section{Circumstellar disks}

Huge progresses were performed thanks to the interferometry (VLTI, CHARA, NPOI) and the multi-wavelengths studies.
The main result is that the disk is rotating in a Keplerian way around its central star, 
see for example \cite[Meilland et al. (2007)]{meilland2007}.
In addition, the thermal structure of the CS disk is also better known (see Tycner contribution, this volume).

\section{Stellar formation and evolution, Gamma Ray Bursts}

What kind of objects is at the origin of Be stars and their fast rotation? Could they be related to the Herbig Ae/Be objects?
Are the fast B rotator Bn stars at the Be star origin, while their number is roughly equal to the number of Be stars in the MW?
What is the IMF of the Be stars? Does it differ from the normal B stars?
It seems that the open cluster density has no effect on the appearance of Be stars (\cite[McSwain \& Gies 2005]{McSwaingies2005},
\cite[Martayan et al. 2010a]{marta2010a}).

\begin{figure}[h!]
% \vspace*{-2.0 cm}
\begin{center}
\includegraphics[width=2.5in]{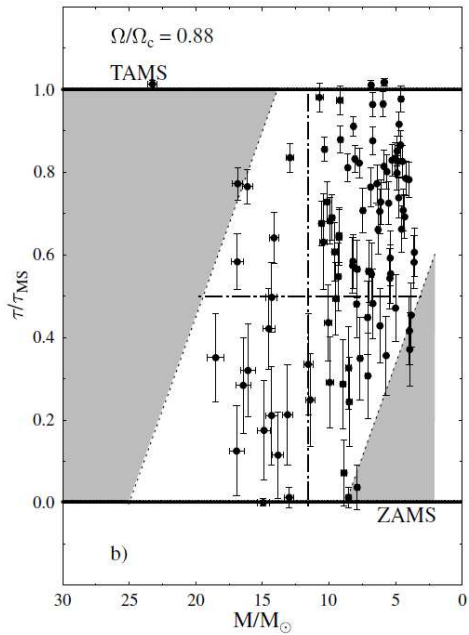} 
% \vspace*{-1.0 cm}
 \caption{Evolutionary diagram in the MS of the Be stars in the MW by \cite[Zorec et al. (2005)]{zorec2005}.}
   \label{fig10}
\end{center}
\end{figure}

Taking into account the fast rotation effects, \cite[Zorec et al. (2005)]{zorec2005} derived the evolutionary status of Be stars in
the MW (see Fig.~\ref{fig10}). Their location/life in the MS strongly depends on their mass and on the evolution 
of the $\Omega/\Omega$$_{c}$ ratio.
In the LMC, the diagram of Be stars appearance is similar than in the MW but in the SMC, at the opposite of the MW, massive Be stars
and Oe stars appear or are still existing in the second part of the MS (\cite[Martayan et al. 2007a]{marta2007a}). 
These differences are also consistent with the $\Omega/\Omega$$_{c}$ evolution at different metallicities.

However, the life of a Be star is not finished at the end of the MS. It seems that the more massive SMC Be and Oe stars could play a
role in the explanation of the type 2 GRBs (also called Long Gamma Ray Bursts). \cite[Martayan et al. (2010b)]{marta2010b} 
were able to reproduce the number and ratio of LGRBs using the populations of SMC massive Be and Oe stars accross 
their stellar evolution.
The type 3 GRBs (without SN counterpart) could be explained in a binary scenario involving Be-binary stars
(\cite[Tutukov \& Fedorova 2007]{tutu2007}).

\section{Binarity}

While 60 to 70\% of O stars are known to be binaries (see \cite[Sana et al. 2009]{sana2009} and his contribution in this volume), 
the question remains opened for the B and Be stars.
 The last study by \cite[Oudmaijer et al. (2010)]{oud2010} using the VLT-NACO AO facility found that:
 29 to 35 $\pm$8\% of the nearby B stars are binaries and that 30 to 33 $\pm$8\% of the nearby Be stars are binaries.
 They scanned the separations larger than 20 AUs with a flux contrast ratio of 10 corresponding to M companions.
Scanning smaller scales needs the interferometry. About 40 \% of all Be stars observed with the VLTI and CHARA 
were found to have a companion.

However, certain type of Be stars are known to be binaries such as the Be-X rays (see for example \cite[Coe et al. 2008]{coe2008}).
The $\gamma$ Cas-like stars have possible companions and have magnetic field activity in their CS environment (\cite[Smith et al. 2006]{smith2006}).

About the Be-binary disruptions scenario when the companion has exploded in SN, one should find that Be stars are runaway stars or see SN
remnants. In both cases, this is not observed for the large majority of Be stars. 
The chemical abundances could help to find a potential previous interaction with a companion. 
In that case, one can expect abundances anomalies.

\section{Conclusion}

To better understand the properties of these stars, it is necessary to do multiwavelength studies and multi-techniques studies
with the use for example of: the VLT-XSHOOTER, IR and X-rays satellites, MOS such as the VLT-FLAMES, the interferometry, 
the AO facilities, and if possible simultaneous observations.

In the future, it will be possible to do similar studies than today but in very far galaxies, to scan other metallicity ranges, etc,
thanks to the ELTs and their instruments but also to ALMA, SKA, etc.

As shown, the Be phenomenon could be not restricted to the late O, B, and early A stars and could
also not be restricted to the MS, specially in lower metallicity environments. 
Some of the questions about them and their CS environment are now solved but some others are still
opened and the definition of Be stars should be revised.

Finally, one can propose this new definition of the Be phenomenon:

{\it This is a star with innate or acquired very fast rotation, which combined to other mechanism such as non-radial pulsations beating
leads to episodic matter ejections creating a CS decretion disk or envelope.}

This definition implies that the stars are not restricted to the MS, not restricted to B-type, and from the CS material the
emission lines come.

\begin{discussion}

\discuss{T. Rivinius}{Comment on the remark by A. Miroshnichenko about the Be binary fraction.
The binary fraction could well be 50 \%. What is really important is that in studies where B and Be
stars are investigated, after having made sure that the same selection effects apply to both, the binary fraction
is found the same for B and Be stars. At least for binary separations wide enough not to destroy the disk
by interaction.}

\discuss{R. Prinja}{Are the fast, high-ionization polar winds of Be star clumps (and perhaps structures)
in a manner akin to massive OB stars? Fundamental differences in wind characteristics of these classes of
hot stars may constrain the pivotal role of for example Fe-bump driven convection 
and the role of disk-wind driving in Be stars. }

\discuss{C. Martayan}{The clumping and the wind structure of Be stars are not yet known. Maybe the current interferometric
programs on Be stars could bring some preliminary informations.} 

\discuss{A. Maeder}{During the cycle Be-B, are luminosity variations observed?}

\discuss{C. Martayan}{Yes, during the life cycle of the CS disk of Be stars, 
some luminosity variations were reported and they follow a sequence 
that is shown by De Wit et al. (2006).}
\end{discussion}

\end{document}